\documentstyle[preprint,aps,prb]{revtex}
\draft

\begin{document}

\title{Low Temperature Properties of Quantum Antiferromagnetic Chains with
       Alternating Spins $S=1$ and $1/2$}

\author{S. Brehmer$^1$, H.-J. Mikeska$^1$, and S. Yamamoto$^{1,2}$}
\address{$^1$Institut~f\"ur~Theoretische~Physik, Universit\"at~Hannover,
         30167~Hannover, Germany\\
         $^2$Department of Physics, Faculty of Science, Osaka University,
         Toyonaka, Osaka 560, Japan} 

\maketitle

\begin{abstract}
        We study the low-temperature properties of $S=1$ and $1/2$ alternating
        spin chains with antiferromagnetic nearest-neighbor exchange
        couplings using analytical techniques as well as a quantum Monte
        Carlo method.
        The spin-wave approach predicts two different low-lying excitations,
        which are gapped and gapless, respectively.
        The structure of low-lying levels is also discussed by perturbation
        theory in the strength of the Ising anisotropy.
        These analytical findings are compared with the results of
        quantum Monte Carlo
        calculations and it turns out that spin-wave theory well
        describes the present system.
        We conclude that the quantum ferrimagnetic chain exhibits 
        both ferromagnetic and antiferromagnetic aspects.
\end{abstract}

\medskip


\section{Introduction}

The low temperature properties of low-dimensional quantum antiferromagnets
have been of great interest for many years, in particular since 
Haldane [\onlinecite{HA}] made the prediction that integer-spin 
and half-odd-integer-spin Heisenberg antiferromagnetic chains should 
behave very differently. Now, with the help of analytical methods and
various numerical approaches, it is well accepted that the
integer-spin chain is massive, whereas the half-odd-integer-spin chain is
massless. This stimulated also several attempts to investigate the
quantum behaviour of chains consisting of two types of spins.
An integrable model of this type was constructed by de Vega and 
Woynarovich [\onlinecite{VW}], which allows us to guess the
essential consequences of chains with spins of different length $S$.
Recently, several authors [\onlinecite{VMN,FFK}] discussed in 
detail such a chain with spins $S=1$ and $1/2$ as the simplest case.
However, these integrable models usually include complicated interactions
and very little is known about the pure Heisenberg model
which is supposed to describe real ferrimagnetic compounds
[\onlinecite{Kahn}].

In the following we therefore study the $S=1$ and $1/2$ Heisenberg
antiferromagnetic chain defined by the hamiltonian (\ref{E:H})
\begin{equation}
   {\cal H}=J\sum_{i=1}^L
           \left[
            (\mbox{\boldmath$S$}_{i} \cdot \mbox{\boldmath$s$}_{i})_\lambda
           +(\mbox{\boldmath$s$}_{i} \cdot \mbox{\boldmath$S$}_{i+1})_\lambda
           \right] \,,
   \label{E:H}
\end{equation}
where
$(\mbox{\boldmath$S$}_{i} \cdot \mbox{\boldmath$s$}_{j})_\lambda
 = \lambda (\, S_i^xs_j^x+S_i^ys_j^y \,)+ S_i^zs_j^z$ with
$\mbox{\boldmath$S$}_{i}$ and $\mbox{\boldmath$s$}_{i}$ being the $S=1$ and
$S=1/2$ spin operators, respectively, and $L$ is the number of unit cells.
We adopt periodic boundary conditions and investigate ground-state 
properties as well as the low-lying excited states spin-wave theory as 
well as series expansions with regard to the Ising anisotropy $\lambda$
and a quantum Monte Carlo (QMC) method.

Basic predictions on the low lying level structure can be made by
applying the Lieb-Mattis-Theorem [\onlinecite{LM}] and the Goldstone-Theorem
[\onlinecite{GS}]. First we consider the ordering of the energy levels.
The {\it A} sublattice is defined by the $S=1$ sites (maximum spin is 
$S_A=L$), the {\it B} sublattice is defined by the spin-1/2 sites (maximum 
spin is $S_A=L$). The Hamiltonian (\ref{E:H}) consists of intersublattice
interactions only, which allows us to apply [\onlinecite{LM}] to our system.
The energy levels order in the following manner:
$$
E(S+1) > E(S) \quad  \mbox{for all}  \quad S \geq {\cal S} \qquad
E(S) > E( {\cal S} ) \quad  \mbox{for}  \quad S < {\cal S}  \qquad
$$ 
Here $\; {\cal S} = | S_A - S_B| \; $ denotes 
the total spin of the ground state
which takes the value  $ \; {\cal S} = L - L/2 = L/2 \;$ in our case.
Therefore the ground-state degeneracy is $L+1$. 
Although the ground states have finite spin ${\cal S}$
the following points are in contrast to the usually ferromagnet:
\begin{itemize}
\item The ground-state degeneracy of the ferromagnet is larger by a
      macroscopic amount.
\item Due to quantum fluctuations the ground state deviates 
	from the ferrimagnetic N\'eel state,
      $| +1 , -1/2 , +1 , -1/2, \ldots \rangle$.
\end{itemize}
Because each of the ground states of the chain breaks the rotational
symmetry of the Hamiltonian, we can apply the Goldstone-Theorem
[\onlinecite{GS}] to predict a gapless excitation. For magnetizations
lower than $L/2$ the ferrimagnet is comparable to a ferromagnet
and therefore this gapless excitation should belong to a
branch of excitations of ferromagnetic character.

The paper is organized as follows: In the second section 
we present the spin-wave approach to calculate the dispersion
relations as well as 
the ground-state energy and the ground-state correlation functions. 
In the following section we study the low-temperature properties
employing a QMC method. The numerical data
will be compared to the results  of the first section and
to perturbation calculations. Conclusions will be given in the final section.

\section{SPIN WAVE THEORY} \label{S:SW}

In this section the spin wave theory (see e.g. [\onlinecite{MA}])
is applied to our system 
in the large $S$ limit. For this purpose the odd sites are
assigned spins $gS$ and the even sites spins $S$.
To discuss the case of our Hamiltonian (\ref{E:H}) we use $g=2$ 
large $S$ limit. We start from the fully ordered state with 
$S^z_{tot}=M=L/2$ and use the following spin operators in the two 
sublattices in the lowest order in $1/S$ :

\begin{eqnarray}
{\rm sublattice \quad A: \;}& &  
	S_{n}^z = gS - a_{n}^{\dagger} a_{n}\,, \qquad 
	S_{n}^+ = \sqrt{2gS} a_{n}\,, \label{transA} \\ 
{\rm sublattice \quad B:\;}& &  s_{n}^z = -S + b_{n}^{\dagger} b_{n} \qquad  
	s_{n}^+ = \sqrt{2S} b^{\dagger}_{n}\,, \label{transB}
\end{eqnarray}

Here $ a_{n}^{\dagger} ,  b^{\dagger}_{n}$ are Bose operators. 
We expect the spin wave theory to give qualitatively correct
results for the following reason: In the classical antiferromagnet 
the two Neel states become disconnected by the 
transformation (\ref{transA},\ref{transB}) since the action of the 
transverse Hamiltonian on one of the two Neel states does not
lead to the other N\'eel state. However, it is known that
domain wall excitations, namely excitations connecting the two
Neel states, are very important for the spin-$1/2$ antiferromagnet.
Spin wave theory is wrong for the spin-$1/2$ antiferromagnet since 
domain wall excitations cannot be taken into account.
In the spin $1$-$1/2$
system the two N\'eel states are disconnected in principle because
they belong to subspaces with different magnetization. Thus a
classical ground state including quantum fluctuations should be 
qualitatively correct.\\
In the limit $S \rightarrow \infty$ the interaction terms in the 
Bose Hamiltonian become negligible and we end up with the following
expression for the Hamiltonian, bilinear in the Bose operators,
\begin{equation}
H_{SW} = -2gS^2JL + 2SJ \sum_k  \left[ 
g b^{\dagger}_{k} b_{k} + a^{\dagger}_{k} a_{k} 
+ \sqrt{g} \cos(k) ( a^{\dagger}_{k} b^{\dagger}_{k} + 
a_{k} b_{k} ) \right]\,, 
\end{equation}
where
$a_n^\dagger=L^{-1/2}\sum_k{\rm e}^{ {\rm i}kn}a_k^\dagger$ and
$b_n^\dagger=L^{-1/2}\sum_k{\rm e}^{-{\rm i}kn}b_k^\dagger$.
The Hamiltonian is straightforwardly diagonalized into
\begin{eqnarray}
	H_{SWT} &=& E_0 + 2SJ \sum_k 
    	\left( \omega_k^- \alpha_k^{\dagger} \alpha_k +
    	\omega_k^+ \beta_k^{\dagger} \beta_k \right) \,, \\ 
   	\omega_k^{\pm} & = & \frac{1}{2} 
	\left( \sqrt{(g+1)^2-4g \cos^2(k)} \pm (g-1) \right)\,, \\
        E_0 &=& JS\sum_k
        \left[
        \sqrt{(1+g)^2-4g\cos^2(k)}-(1+g)
        \right]
        -gJS^2N \,,
\end{eqnarray}
where the eigenvectors are determined by
\begin{eqnarray}
	\alpha^{\dagger}_{k} = \cosh(\eta_k) a^{\dagger}_{k} + 
   	\sinh(\eta_k) b_k \,, &    &
   	\beta^{\dagger}_k = \sinh(\eta_k) a_{k} + 
	\cosh(\eta_k) b^{\dagger}_k\,, \\
    	\tanh(2 \eta_k) &=&  \frac{2 \sqrt{g}}{1+g} \cos(k)\,.
        \label{eta}
\end{eqnarray}
The results of the spin wave theory contain both ferromagnetic
and antiferromagnetic aspects:
In the ferromagnetic branch ($\omega^-_k$) we obtain a gapless spin wave 
with $M=L/2 -1$. For small values of the wave vector
the dispersion is $\; \omega^-_k = g/(g-1) \; k^2 + \ldots \;$ 
The quadratic behaviour with wave vector $k$ indicates 
the ferromagnetic character of this mode.\\
The antiferromagnetic spin wave with $M=L/2 +1$ is gapped.
The magnitude of the gap is exactly $J$ when we put $g=2, S=\frac{1}{2}$.
This result will be compared to the result of the
QMC calculation to be presented in the next section.
If we chose the spins such that the spin lengths in the different
sublattices are equal, i.e. $g \rightarrow 1$
the gap of the antiferromagnetic branch vanishes.
Another interesting feature is the ground-state sublattice magnetization 
as defined by
\begin{eqnarray}
\langle M_A \rangle_{g.s.}  &=& L  - \sum_{n=1}^L a^{\dagger}_{n} a_{n}
= L ( 1 - \tau )  \\ \tau &=& \frac{1}{\pi} \int_0^{\infty} dk 
\sinh^2( \eta_k) \approx 0.305 \;.
\end{eqnarray}
Notice that the one dimensional ferrimagnet has a finite spin
reduction $\tau$. This is due to 
the factor $2\sqrt{2}/3 < 1$, obtained for $g=2$ in relation (\ref{eta}). 
In the usual
antiferromagnet the corresponding factor is $1$ and leads to a 
diverging spin reduction (in 1 D). \\
The ground-state energy per unit cell in the spin wave approach 
is obtained as
\begin{equation}
	E_0/J  =  - \frac{5}{2} L + \frac{L}{2 \pi} \int_0^{\pi} dk
   	\sqrt{9-8 \cos^2(k)} \approx -1.4365L 
	\qquad( S= \frac{1}{2} \; , \; g=2 ). 
\end{equation}

Further we studied the ground-state correlation functions
between two sites with $S=1$ for the ground state with magnetization $L/2$
and obtained the following results:
\begin{eqnarray}
   \langle S_{n}^z S_{n+r}^z \rangle
   &=& (1 - \tau)^2 + f(r) \\
   \langle S_{n}^+ S_{n+r}^- \rangle
   & \propto & \sqrt{ f(r) } \\ 
   f(r) &=& \frac{1}{L^2} \sum_{k,q}
   e^{{\rm i} 2qr} \sinh^2(\eta_k) \cosh^2(\eta_{k+q}) \label{fr}
\end{eqnarray}
The asymptotic behaviour for the function $f(r)$ 
is calculated by taking the continuum limit for the sum in eq.(\ref{fr}).
We observe asymptotical exponential decay with
a correlation length $\xi$, $\xi^{-1}=2 \ln(2)$.
So for the
intrasublattice distance $r \geq 2$ correlations  have decayed and
the square of the spin reduction $ (1 - \tau)^2 $ remains.
From comparison with the QMC results it will be shown later this is a 
qualitatively correct picture.

As conclusion of this section we discuss the case
of strong alternation, a limit which naturally reproduces the scenario 
as given above. If we introduce the interaction 
strength $J$ between sites $(2n)$ and $(2n+1)$
and  $\delta J$ between sites $(2n-1)$ and $(2n)$ the Lieb-Mattis-Theorem
and the Goldstone-Theorem still hold. The $M=L/2$
ground state in the dimerized limit $\delta=0$ can be written down as
\begin{eqnarray*}
| 0 \rangle = | D_+ \rangle_1  | D_+ \rangle_2  & \ldots & | D_+ \rangle_L \\
( \vec{S}_{n}+ \vec{s}_{n})^2 | D_+ \rangle_n = \frac{1}{2}
\left( \frac{1}{2} +1 \right) | D_+ \rangle_n & \qquad &
( S^z_{n}+ s^z_{n}) | D_+ \rangle_n = \frac{1}{2} | D_+ \rangle_n \\
 | D_+ \rangle = \frac{1}{\sqrt{3}} ( | 0 , +1/2 \rangle   - 
\sqrt{2} | +1 , -1/2 \rangle
\end{eqnarray*}
For values  $\delta \ll 1$ the system behaves like a
$S=\frac{1}{2}$ ferromagnet with the dublets ($D_+, D_-$) being the
effective spin $S_{eff}=\frac{1}{2}$. A first order perturbation 
calculation in
the alternation parameter $\delta$ leads to a ground-state energy
of $E_0= -L(1+1/9) +   {\cal O}( \delta^2)$. Now a ferromagnetic
spin wave can be constructed within the dublet subspace
\begin{equation}
   | q \rangle = \frac{1}{\sqrt{L}}
   \sum_{n=1}^L e^{{\rm i}qn} | D_+ \rangle_1 \ldots  
   | D_- \rangle_n \ldots | D_+ \rangle_L 
\end{equation}
The dispersion up to the first order in $\delta$ is $\omega (q) =
4/9 \lambda (1- \cos(q)) + {\cal O}( \delta^2)$. It is gapless 
and proportional to $q^2$ for small wavevectors as one would expect.\\
In addition to the ferromagnetic excitations we can construct
antiferromagnetic spin waves, i.e. spin waves with $S_{tot}^z= L/2+1$.
Again we discuss the case of strong alternation and end up with 
\begin{equation}
| q \rangle = \frac{1}{\sqrt{L}} \sum_{n=1}^L e^{iqn} | D_+ \rangle_1 \ldots  
| Q_3 \rangle_n \ldots | D_+ \rangle_L 
\end{equation}
Here $ \; | Q_3 \rangle \; $ denotes the quartet with magnetization
3/2 . The first order dispersion is $\omega (q) = 3/2 + \lambda
( 7/18 -2/3 \cos(q) ) +  {\cal O}( \delta^2)$  and is gapped.

\section{NUMERICAL RESULTS}

\subsection{Brief Account on the Numerical Procedure}

In the following  we employ a quantum Monte Carlo method based on the 
Suzuki-Trotter
decomposition [\onlinecite{Suzu}] of checker-board type [\onlinecite{HSSB}].
Raw data are taken for a set of Trotter numbers $n$ and are extrapolated 
to the
$n\rightarrow\infty$ limit with the parabolic fitting formula.
We carry out all the calculations in certain subspaces with a fixed value of
the total magnetization.
Since we treat the chains with periodic boundary conditions, not only the
Monte Carlo flips of local type but also the global flips along the chain
direction are taken into the numerical procedure.
On the other hand, global flips along the Trotter direction, which mean
fluctuations of the total magnetization, are not included in order 
to well describe the ground-state properties.
The quantum Monte Carlo algorithm to update the spin configuration is detailed
elsewhere [\onlinecite{Yama}].
We have confirmed that almost the same results are obtained at two
temperatures, $k_{\rm B}T/J=0.04,\ 0.02$, and thus we regard these
temperatures as low enough to successfully extract the lowest-energy-state
properties.
Here we show the data taken at $k_{\rm B}T/J=0.02$.
The data precision is almost four digits for the energy, and two digits for
the spin correlations.

\subsection{Low-Energy Structure}

 In Fig.\ref{generg}
  we plot the lowest energies per unit cell in the subspaces with
$M=\sum_i(S_i^z+s_i^z)=L/2,\, L/2-1$ as a function of $L$.
The coincidence between both sets of data is nothing but the numerical
demonstration of the above mentioned Lieb-Mattis theorem, that is, the
$L/2$-multiplet structure of the ground state.
The rapid convergence into the long-chain limit suggests a rather small
correlation length in this system, which will be actually observed in the
following.
Within the present numerical precision, the ground-state energy in the
thermodynamic limit is estimated as $E_{\rm G}=-1.455\pm 0.001$.
As we have observed in Fig.\ref{generg}, even the 
ground-state energy of the $L=16$
chain is already close to the thermodynamic-limit value. Thus we have plotted
the quantum Monte Carlo data for $L=16$.

Although the quantum Monte Carlo data as presented above are already 
conclusive in themselves, we present an additional argument based on
perturbation theory in the Ising anisotropy $\lambda$. 
We compare in Fig.\ref{compgs} the quantum Monte-Carlo 
estimated ground-state energies and
the corresponding perturbation-theory result
\begin{equation}
   -E_{\rm G}/L = 1 + \frac{ \lambda^2}{2} - \frac{ \lambda^4}{48}
	- 0.05136 \lambda^6 + 0.02809 \lambda^8 +{\cal O}( \lambda^{10})\,,
\end{equation}
as a function of $\lambda$. We find a fairly good
agreement between the eighth-order perturbation result 
and the QMC calculation.
What should be emphasized is that fourth order calculation shows good
agreement with the correct result. This fact gives us an idea on
the spin configuration in the $M=L/2$ ground state.
All the fluctuations introduced within the fourth order are
essentially classified into the following three types:
\begin{eqnarray*}
   {\rm (a)\ \ 2}\mbox{-}{\rm sites\ fluctuation}&:&
   1 , -1/2 , 1 , -1/2 , \; \underbrace{ 0 , 1/2 } \; ,
   1 , -1/2 , 1 , -1/2 , \cdots \\
   {\rm (b)\ \ 3}\mbox{-}{\rm sites\ fluctuation}&:&
   1 , -1/2 , 1 , \; \underbrace{ 1/2 , -1 , 1/2 } \; ,
   1 , -1/2 , 1 , -1/2 , \cdots \\ 
   {\rm (c)\ \ 4}\mbox{-}{\rm sites\ fluctuation}&:&
   1 , -1/2 , \; \underbrace{ 0 , 1/2 , 0 , 1/2 } \; ,
   1 , -1/2 , 1 , -1/2 , \cdots
\end{eqnarray*}
We note that the formation energy of the defect (c) is not twice as much as
that of (a) and therefore they should be distinguished.
Based on these fluctuations we are led to discuss microscopic
quantum fluctuations at the isotropic point $(\lambda=1)$.
We present in Fig.\ref{snap} a QMC snapshot from which we can extract the
image of quantum fluctuations to a certain extent.
Here we show the snapshot at $k_{\rm B}T/J=0.02$ for $\lambda=1.0$, $L=32$,
and $n=20$, where the horizontal and vertical lines denote the chain and
the Trotter directions corresponding to space and time, respectively.
We find out everywhere local defects breaking the N\'eel Order, 
whereas they are
all identified with the above-mentioned fluctuations.
Thus we expect the fourth order calculation to well describe the
the $M=L/2$ ground state.

Next we discuss the lowest excited states 
for the ferromagnetic and the antiferromagnetic branches. 
In the subspace with $M=L/2-1$ we construct a magnon for $\lambda=0$
by flipping a spin-1: $S^z=1 \rightarrow S^z=0$. Expanding this state
and taking the limit $k \rightarrow 0$ we obtain the lowest mode
in this subspace. Such an expansion up to the fourth order is
compared with $E_{G}(L/2-1) - E_{G}(L/2)$ from the QMC calculation 
in Fig.\ref{compex}. The perturbation result gives reasonable results up to
$\lambda \approx 0.8$. A comparison between them suggests that the
$k=0$ magnon condenses into the ground state with $M=L/2-1$
at the isotropic point.\\
The antiferromagnetic magnon is constructed by flipping a spin-1/2
up. Again the perturbation result and the QMC data for 
$E_{G}(L/2+1) - E_{G}(L/2)$ are illustrated in Fig.\ref{compex}. Here the
validity for the fourth order expansion only reaches up to 
$\lambda \approx 0.4$.
From the QMC calculation we obtain a gap of 
$\Delta/J = 1.767 \pm 0.003 $. Here somewhat larger uncertainty,
rather than one for the ground-state energy, mainly comes from the Monte
Carlo estimate in the subspace of $M=L/2+1$.  
This value is much bigger than the prediction of the spin wave
theory $\Delta = J$. 

We conclude this section by pointing out that the quantum behaviour 
of the $1$-$1/2$ system results in an enhancement of the gap. 
From Fig.\ref{compex} we 
observe that the pure Ising energy of $2J$ is just lowered by
a small amount when moving to the isotropic point. 

\subsection{Spin Correlations}

 We show in Fig.\ref{SS} the spin correlation functions
between the spins of the same type 
for the lowest-energy states in the subspaces with
$M=L/2-1,L/2,L/2+1$.
We note that the self correlation of spin-$1$ deviates from $2/3$ because
of the multiplet structure of the ground state.
In comparison with the spin correlations for $M=L/2$, those for $M=L/2-1$ are
significantly reduced rather than those for $M=L/2+1$ in the case of
spin-$1$, and vice versa in the case of spin-$1/2$.
This is well understood considering that
the ferromagnetic and the antiferromagnetic magnons, which exist in
the subspaces with $M=L/2-1$ and $M=L/2+1$, originate from the spin flips
in the $S=1$ and the $S=\frac{1}{2}$ sublattices, respectively.
Those excitations are
expected to reduce the ferromagnetic correlations between
spins with $S=1$ and spins with $S=\frac{1}{2}$.
We note that in the thermodynamic limit, both spin correlations for
$M=L/2-1$ and $M=L/2+1$ should coincide with those for $M=L/2$.
Nevertheless, Fig.\ref{SS} 
is still useful because it suggest to a certain extent
the thermodynamic-limit spin correlations in the subspaces with magnons of
finite density.

In the above sense, now let us concentrate ourselves on the subspace with
$L/2$.
We have already confirmed that the spin wave theory gives the 
asymptotic exponential
decay of the spin correlations, where the correlation length is estimated to
be less than unity.
We here observe so rapid decay of the correlations that an estimate of the
correlation length is beyond the present numerical precision.
However, careful observation of Fig.\ref{SS} 
shows us that the correlations between
spins $S=\frac{1}{2}$ are a little bit less rapid than those between 
spins $S=1$.
Spin wave theory cannot reproduce this feature because the decay
is determined by the function
$f(r)$ in eq.(\ref{fr}) for both sublattices.
We further point out that the Monte Carlo calculation 
gives the spin reduction $\tau$
as $\tau\simeq 0.21$, which is somewhat smaller than the spin-wave-theory
result.
Therefore the spin-wave-theory cannot quantitatively
describe the quantum fluctuations.

\section{conclusion}
We have calculated ground-state properties and low-lying excited states
for an  alternating ferrimagnetic spin chain with spins $S=1$ and 
$\frac{1}{2}$. The ground
state is a spin $S=L/2$ multiplet. The model consist of a ferromagnetic
and an antiferromagnetic branch corresponding to  magnitizations $M < L/2$
and $M > L/2$, respectivly. The ferromagnetic branch has gapless
excitations with dispersions $\omega \propto k^2 \; , \; k \ll \pi$
according to spin wave theory. The antiferromagnetic branch 
with $M=L/2+1$ shows a gapped spin wave with 
$\Delta /J=1.767 \pm 0.003$.
Both branches have longitudinal correlation functions consisting of a
constant (square of spin reduction) plus strong exponential decay
and therefore both branches show long range order.
The manifestation of quantum behaviour shows up in the following
points
\begin{itemize}
\item the gap for the $M=L/2+1$ excitation is enhanced compared
	to the spin wave theory.
\item the $L/2$ ground state deviates from the 
	N\'eel state owing to quantum fluctuations.
\item there is an indication for the correlation length to be
	larger in the $S=1/2$ than $S=1$ sublattice.
\end{itemize}
In order to study the mechanism for the gap formation in detail
we have to investigate the perturbation theory to higher order.
This is under investigation 
as well as the construction of matrix product states as
variational ground state for $M=L/2$ [\onlinecite{KMY}].
The construction of these matrices can be based on the fact that
unit cells with magnetization $-3/2$ show up rather rarely
in Fig.\ref{snap}. Therefore the system exhibits a tendency
to weak ferromagnetism similar to the one discussed by Niggemann and
Zittartz [\onlinecite{NZ}] in $S=3/2$ chains with matrix product 
ground states.

\section*{acknowledgements}

This work was supported by the German Federal Minister of Research and
Technology (BMBF) under contract number 03-MI4HAN-8.


\newpage
\begin{figure}
\caption{Size dependences of the ground-state energy in
	the subspaces with $M=L/2$ and $M=L/2-1$}
\label{generg}
 \end{figure}
\begin{figure}
\caption{Dependences of the ground-state energy on the Ising
anisotropy $\lambda$ obtained by a QMC method ($\Diamond$), 
and 4-th-order (dotted line) 
and 8-th-order (solid line) perturbation theory.}
\label{compgs}
\end{figure}
\begin{figure}
\caption{A snapshot of the transformed two-dimensional Ising system,
	where the horizontal and the vertical lines denote the chain and
	the Trotter directions corresponding to space and time, respectivly,
	and `\#', `+', `0', `-', and `=' denote the spin projections
	$+1$, $+1/2$, $0$, $-1/2$, $-1$ $\,$.}	 
\label{snap}
\end{figure}
\begin{figure}
\caption{Dependences of the excitation gaps $E_G(L/2-1)-E_G(L/2)$ 
	($\bigcirc$) and
	$E_G(L/2+1)-E_G(L/2)$ ($\Diamond$) on the Ising anisotropy obtained 
	by a QMC method, where the results within the 4-th-order 
	perturbation treatment are also shown (solid lines). }
\label{compex}
\end{figure}
\begin{figure}
\caption{The groud-state spin correlations between spins $S=1$ (a) and between
	 spins $S=1/2$ (b) in the subspaces of $M=L/2,L/2-1$ and $L/2+1$
	for $L=16$.}
\label{SS}
\end{figure}

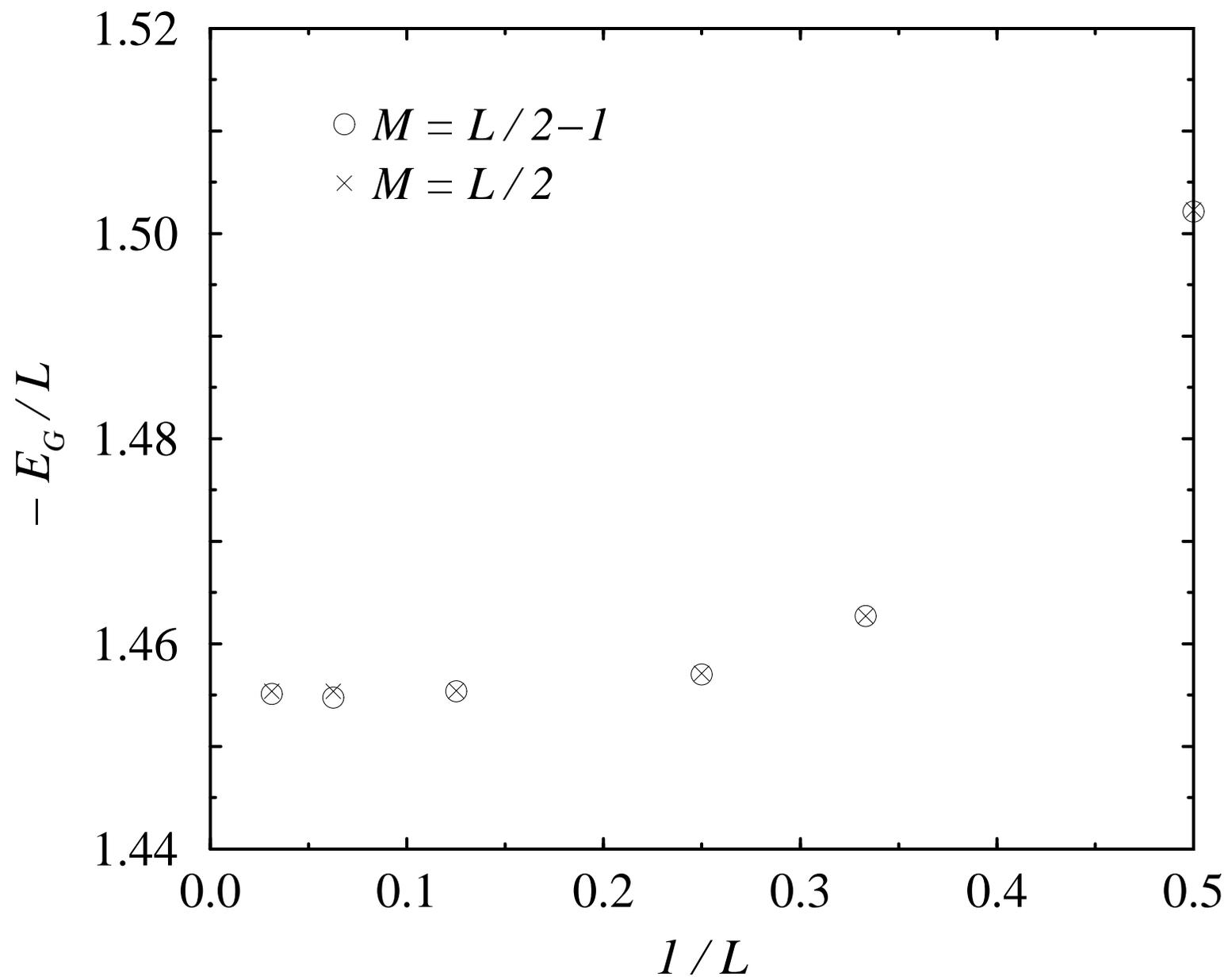
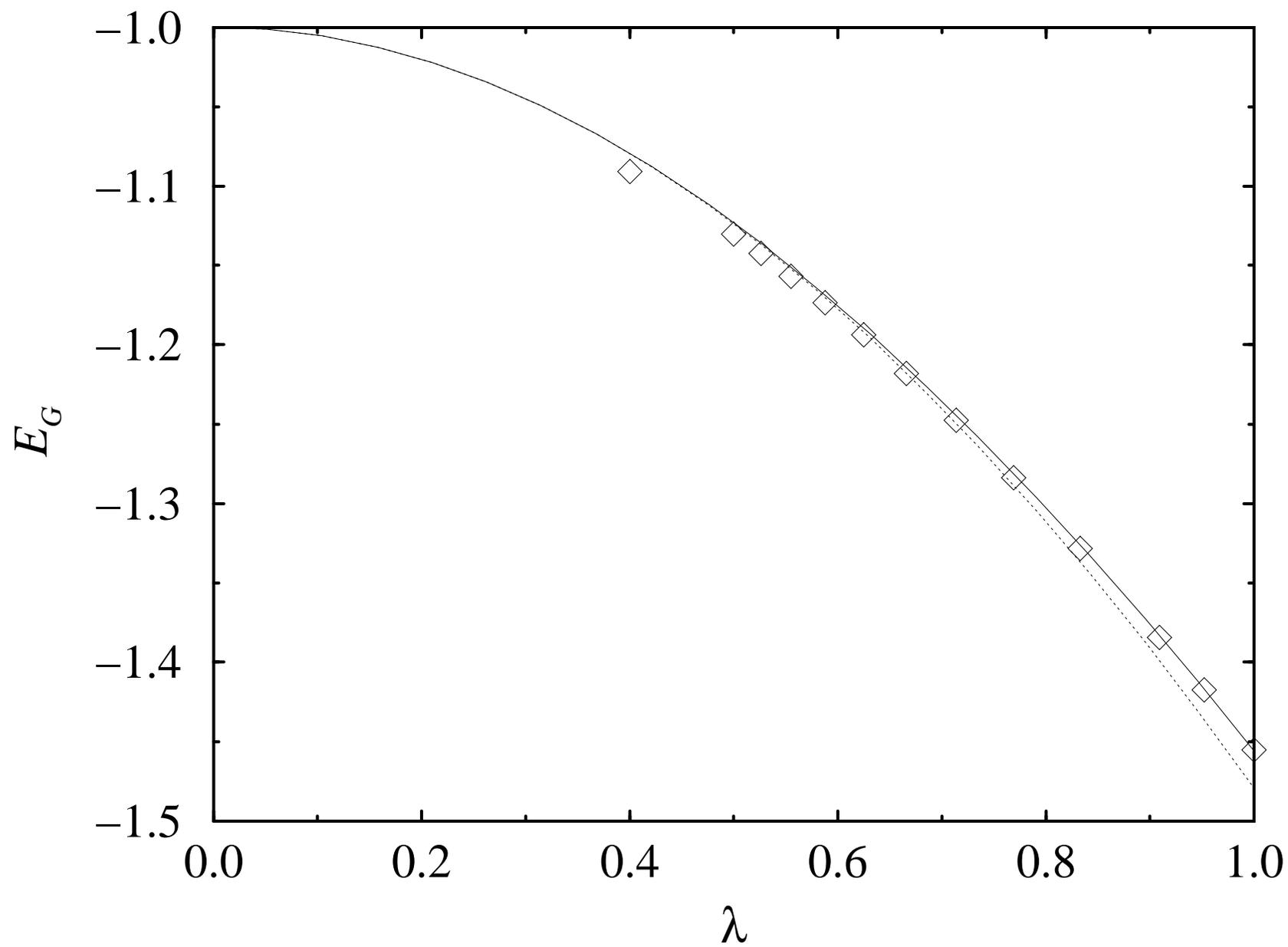
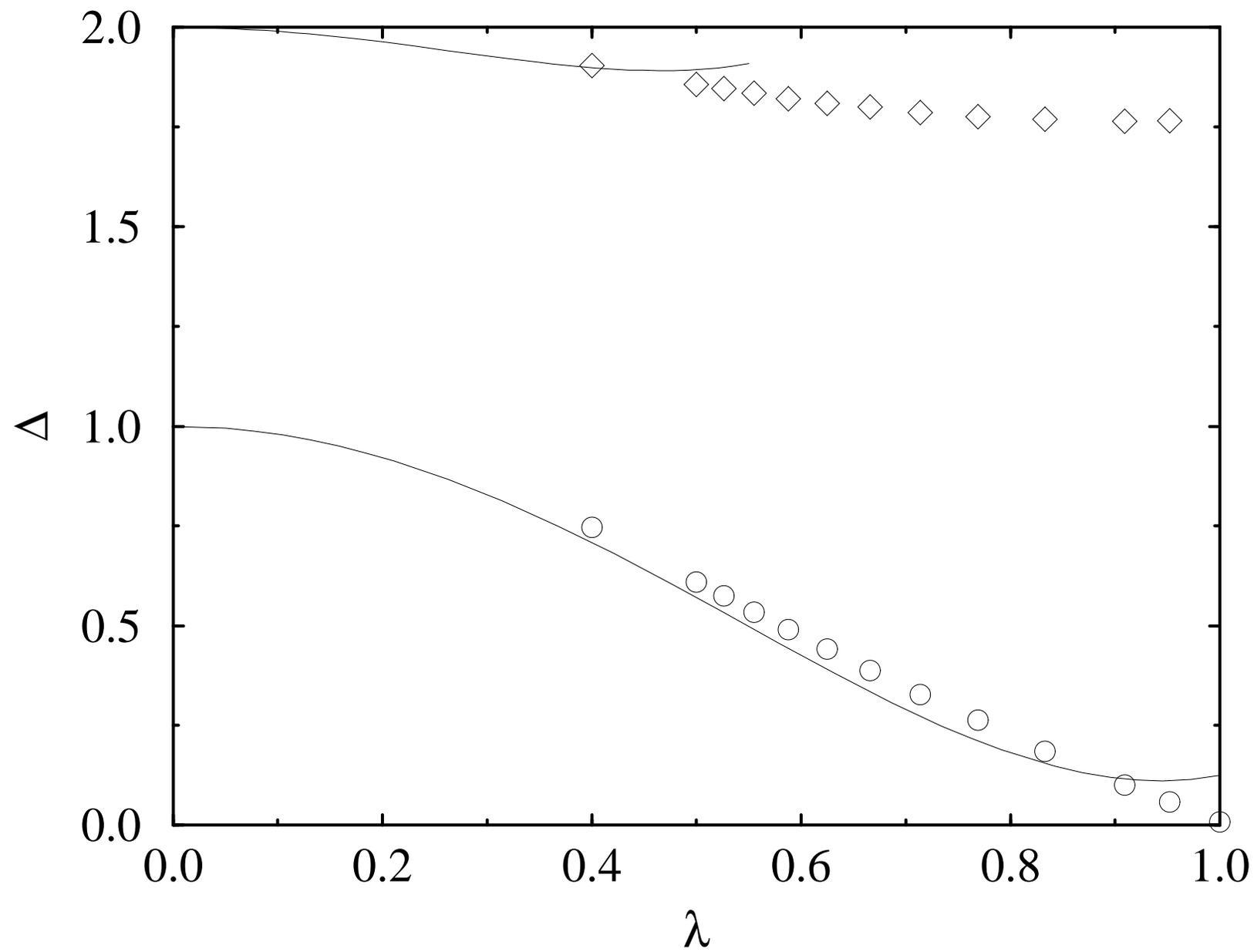
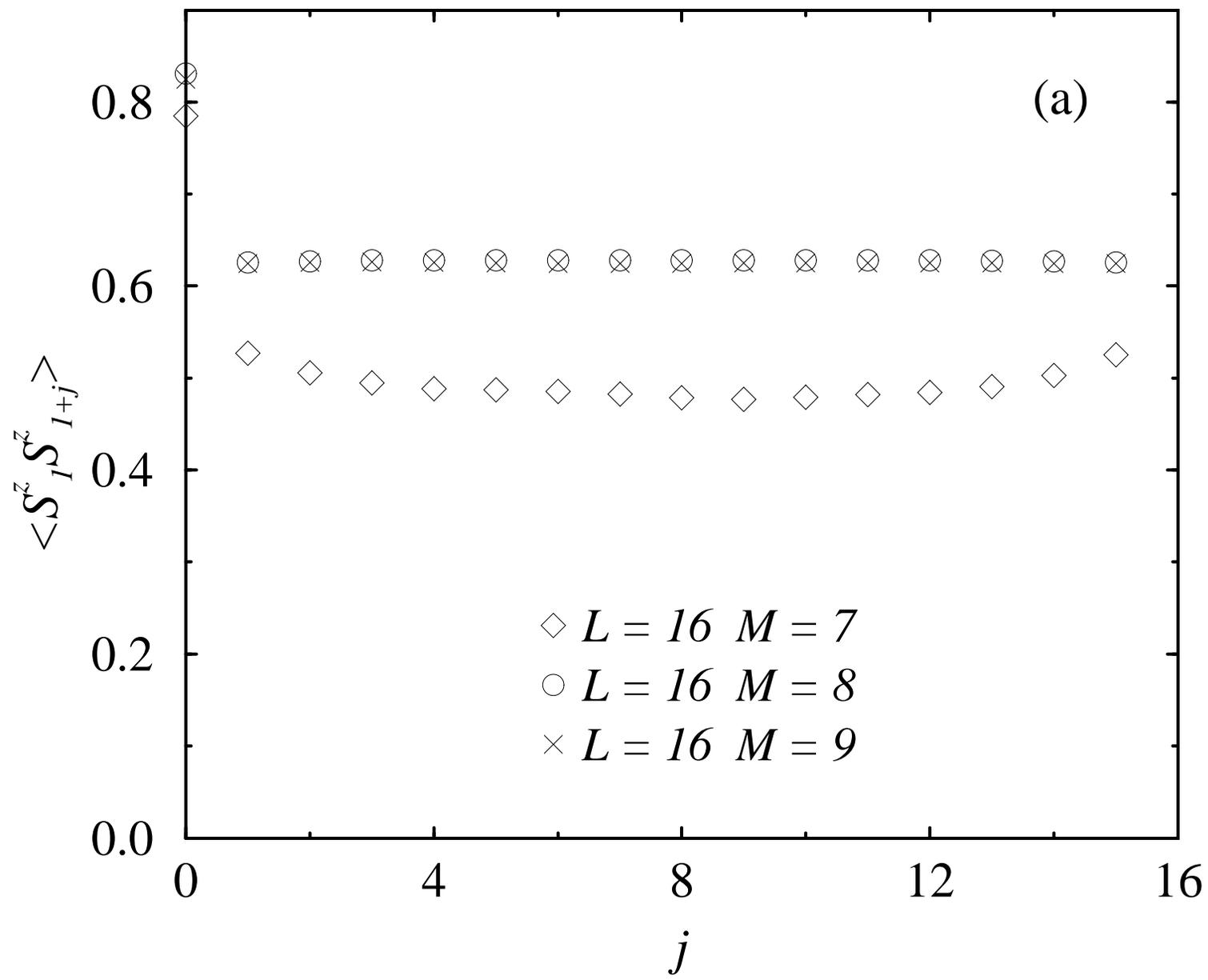
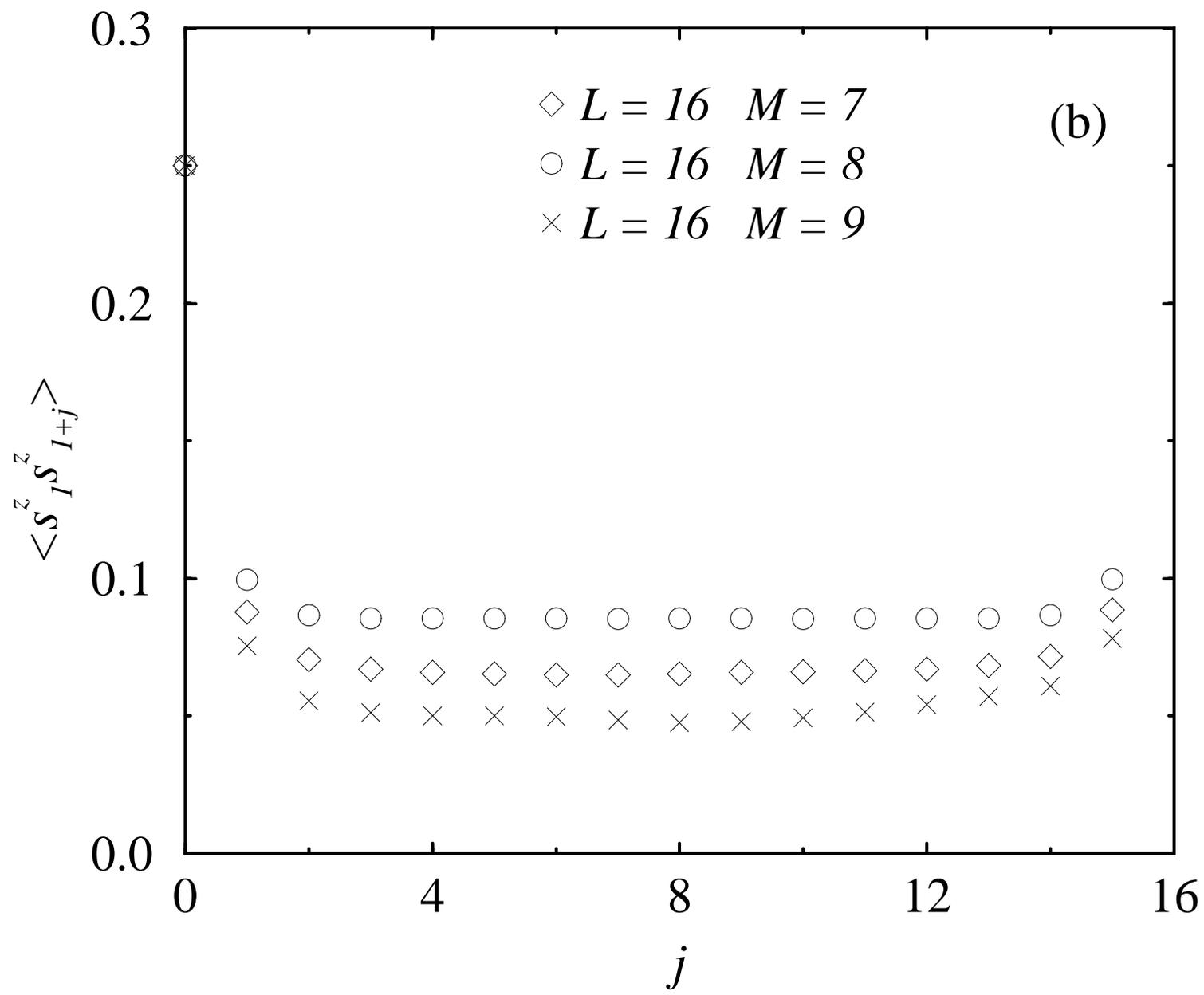
\end{document}